%% file: paper.tex
\documentclass[journal]{vgtc}                     

\onlineid{0}

\vgtccategory{Research}

\title{What Do Visualization Instructors Want Students to Learn? Introducing a Concept Inventory for Visualization Design}

\author{%
  \authororcid{Medina Lamkin}{0000-0002-4366-7994}, \authororcid{Heer Patel}{0009-0003-8958-243X},
  \authororcid{Sayamindu Dasgupta}{0000-0001-6083-2114}, and 
  \authororcid{Leilani Battle}{0000-0003-3870-636X}
}

\authorfooter{
  \item
  	Medina Lamkin, Heer Patel, and Leilani Battle are with the Paul G. Allen School of Computer Science and Engineering at the University of Washington.
  	E-mail: \{mlamkin\,$|$\,heerpate,\,$|$\,leibatt\}@cs.washington.edu\,.
  \item
  	Sayamindu Dasgupta is with the Department of Human Centered Design and Engineering at the University of Washington
  	E-mail: sdg1@uw.edu.
}

\abstract{\input{content/0-abstract}

}


\teaser{
    \centering
    \includegraphics[alt={An image showing the changes made to question 1 between version 1 and version 7. The bar chart showing the 2026 GDPs of USA versus China has been updated to a higher resolution version made with a different visualization toolkit. Both versions of the bar chart feature a truncated y-axis, different colors for the bars, no color legend, and no currency for the GDP. The question in version 1 states “Select only one of the following points to offer as feedback to the creator that would most critically improve this visualization so that a person can accurately compare the GDP of the two economies. ” Version 7 includes a visualization goal of “Help the viewer understand the GDP ration of the USA versus China” and bolds the phrase “most critically improve” to emphasize the goal. For the answer options, rationales have been added to test understanding. “Y-axis should not be truncated” becomes “The y-axis should not be truncated as this causes an overestimation of the difference.” “Color coding is unnecessary” becomes “Color coding is unnecessary  as no additional information is being encoded.” “The color legend is missing” becomes “The color legend is missing, making it unclear if there is additional information being encoded.” “The GDP needs a currency” becomes “The GDP needs a currency in order to get an accurate understanding of the values being compared.”}, width=\textwidth]{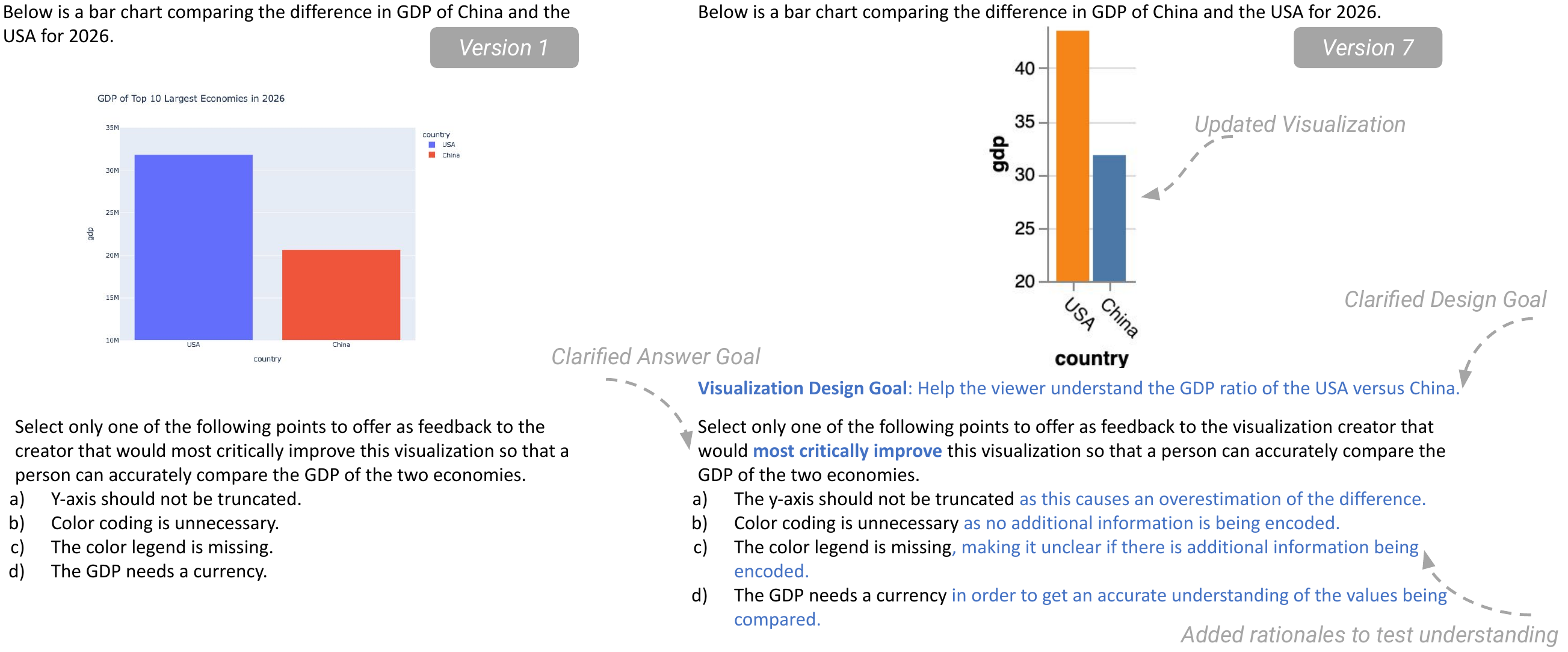}
    \caption{An example question from our \emph{concept inventory}, a 14-question assessment reflecting the breadth of skills that visualization instructors seek to teach their students.
    This question considers a student's ability to detect and mitigate a deceptive visualization tactic: truncating the y-axis.
    We iteratively refined the concept inventory in response to instructor feedback; for example, by adding explicit visualization design goals to guide student evaluation of the solutions and including plausible rationales for each answer to test common misconceptions in visualization design. The rationale behind the design of each answer is shown in \autoref{fig:distractorsEx1}.}
    \label{fig:example1-iteration}
}

\graphicspath{{figs/}{figures/}{pictures/}{images/}{./}} 

\usepackage{tabu}                      
\usepackage{booktabs}                  
\usepackage{lipsum}                    
\usepackage{mwe}                       
\usepackage{ccicons}                   
\usepackage{svg}

\usepackage{mathptmx}         
\usepackage{comment}

\usepackage{multirow}
\usepackage{graphicx}
\usepackage[table,xcdraw]{xcolor}
\usepackage{amssymb}

\newcommand{\dgone}[0]{{{\bf{}DG1}}}
\newcommand{\dgtwo}[0]{{{\bf{}DG2}}}
\newcommand{\dgthree}[0]{{{\bf{}DG3}}}

\begin{document}



\maketitle

\input{content/1-introduction}

\input{content/2-relatedwork}

\input{content/3-methods}
\input{content/4-analysis}

\input{content/5-conceptinventory}

\input{content/6-discussion}

\input{content/7-conclusion}

\section*{Supplemental Materials}
Supplemental materials and instructions for accessing the concept inventory are available on OSF: \href{https://osf.io/dma3f/}{\url{https://osf.io/dma3f/}}.


\acknowledgments{%
	This work was supported in part by the National Science Foundation (Awards \# 2402718, \# 2514565, \# 2141506, \# 2313998). The authors wish to thank all the visualization instructors who gave us permission to use their syllabi, including Andrew McNutt, Bei Wang, Belén Saldías, Chaoli Wang, Charles Perin, Chris Johnson, Cindy Xiong Bearfield, David Laidlaw, Enrico Bertini, Evan Peck, Fateme Rajabiyazdi, Fumeng Yang, Hanspeter Pfister, Jeffrey Heer, Jieqiong Zhao, Joshua Levine, Keke Wu, Klaus Mueller, Marti Hearst, Matt Brehmer, Matthew Beattie, Murtaza Ali, Paul Rosen, Rebecca Williams, Tamara Munzner, Wesley Willett, Yu Fu, and Zhu-Tian Chen. We would also like to thank the professors who provided us with feedback on the concept inventory.%
}

\bibliographystyle{abbrv-doi-hyperref}

\bibliography{references}

\end{document}

%% file: content/0-abstract.tex
\label{abstract}The term ``visualization design'' encompasses multiple concepts and skills that go well beyond current assessments of graphical perception and visualization literacy. 
In the context of education, what exactly should a student be able to do if they ``know'' visualization design? To answer this question, we draw on existing methodology from the field of education to propose a \emph{concept inventory} for visualization design, i.e., a theoretical model capturing the most important concepts and skills commonly associated with visualization design. We initially draft the concept inventory using a qualitative analysis of course objectives from visualization course syllabi. Then, we iteratively refine the concept inventory by soliciting feedback from instructors through semi-structured interviews. Based on our experiences in developing the concept inventory, we reflect on open questions and future research directions in visualization education, such as developing assessments for visualization design (similar to those for visualization literacy) and providing automated assistance for learning and teaching visualization design. Our supplemental materials are available at \href{https://osf.io/dma3f/}{\url{https://osf.io/dma3f/}.}

%% file: content/1-introduction.tex
\section{Introduction}
\label{introduction}

Data visualization courses encompass many skills that students must learn~\cite{Bach2023VisEdChallenges,borner2019data}, making it difficult to understand what core concepts and skills are associated specifically with visualization design. For example, how does a particular instructor discern whether a student truly knows visualization design? Without a clear answer to this question, we argue that it is difficult if not impossible to design effective support tools, AI assistants, and even course materials that enhance a student's knowledge of and ability to apply visualization design.
In this paper, we seek to answer two key research questions:
\begin{itemize}
    \item What are the core topics and skills that visualization instructors want their students to learn through their data visualization class?
    \item How can we gauge the breadth of visualization design principles and methods that a student has learned?
\end{itemize}

While we observe notable research on measuring visualization literacy (e.g., \cite{minivlat,vlat,ge2026autoethnography,ge2025avec,ge2023calvi,borner2019data}), understanding what students learn in visualization courses~\cite{whatStudentsLearnInVisHedayati}, applying learning objectives to visualization (e.g., \cite{adar2021communicative,lee-robbins2022learning,lee-robbins2023affective,burns2020how}), and tools to enhance student learning of data visualization (e.g., \cite{hull2023visgraderautomaticgradingd3,mahadalkar2023dvc,kim2024chatgpt}), we do not see research investigating what concepts and skills visualization instructors consider essential to learning visualization design.
For example, automatic grading tools, such as VisGrader are designed to help instructors conduct a holistic evaluation of assignment code from hundreds of students~\cite{hull2023visgraderautomaticgradingd3}, but they do not clarify what visualization design principles and skills must be leveraged to complete the graded assignments.  Visualization literacy evaluation tools, including interpretation-based assessments like Mini-VLAT~\cite{minivlat}, VLAT~\cite{vlat} and CALVI~\cite{ge2023calvi} and construction-based assessments like AVEC~\cite{ge2025avec}, can be used to measure students' visualization literacy levels as they improve throughout a data visualization course~\cite{whatStudentsLearnInVisHedayati}.
While current assessments emphasize in-depth assessment of specific skills, we observe a lack of research on the \emph{breadth} of concepts and skills 
that instructors consider essential to visualization design, which has been observed in part through prior work~\cite{whatStudentsLearnInVisHedayati}.
Other research identifies common mistakes and pitfalls and suggests changes in visualization education to prevent them~\cite{visPitfallsinSciPubs, fox2020surfacingmisconceptionsvisualizationcritique}.
We seek to extend these ideas by coalescing them into a precise theory summarizing what visualization instructors consider essential to learning visualization design.

To answer our research questions and address current gaps in the visualization literature, we draw on existing theory methods from education research. Specifically, education researchers regularly introduce a specific type of model, a \emph{concept inventory}, to summarize the essential skills and concepts students should know as well as common misconceptions they should avoid within a pre-defined educational scope~\cite{taylor2014computer}. 
Concept inventories have been used by instructors in multiple disciplines to encapsulate the abilities that students should master for a given course topic,
to measure where students may need more support, and to evaluate the effectiveness of a course in addressing areas of weakness \cite{MurtazaCIReview, forceCIHestenes, graphicalCommunicationCISteinhauer}; examples include concept inventories in physics (e.g., the Force Concept Inventory, focusing on Newtonian concepts~\cite{forceCIHestenes}), biology~\cite{10.1641/B581111} (e.g., for natural selection~\cite{https://doi.org/10.1002/tea.10053}), engineering~\cite{fakiyesi2024scoping} (e.g., for fluid mechanics~\cite{martin2003development}), and computer science (e.g., introductory programming~\cite{TewCSCI, caceffoCSCI,programmingCIreplication}).
Concept inventories are used primarily by researchers and instructors. Although students may complete a concept inventory for assessment purposes, the results are typically used to grade the course or pedagogy, not the students~\cite{Sands2018CIsForUnderstanding}.

In line with how concept inventories are typically designed \cite{10.1063/1.2508680}, our concept inventory serves two purposes. First, it synthesizes the \emph{most common topics and skills} that instructors emphasize when describing visualization design within their course learning objectives. Second, it provides \emph{demonstrative example questions} showing how instructors could assess these topics and skills, providing a more precise theory behind what instructors intend for students to learn (and \emph{not} learn) about visualization design. Traditionally, concept inventory questions are designed to be multiple choice to enable scalable assessment, a convention we also follow. An example question is shown in \autoref{fig:example1-iteration}.

\begin{figure}
    \centering
    \includegraphics[alt={An image showing the methods applied, surveying instructors and collecting syllabi, extracting learning objectives, clustering and coding learning objectives, creating assessments  questions, and interviewing instructors for feedback.}, width=0.75\linewidth]{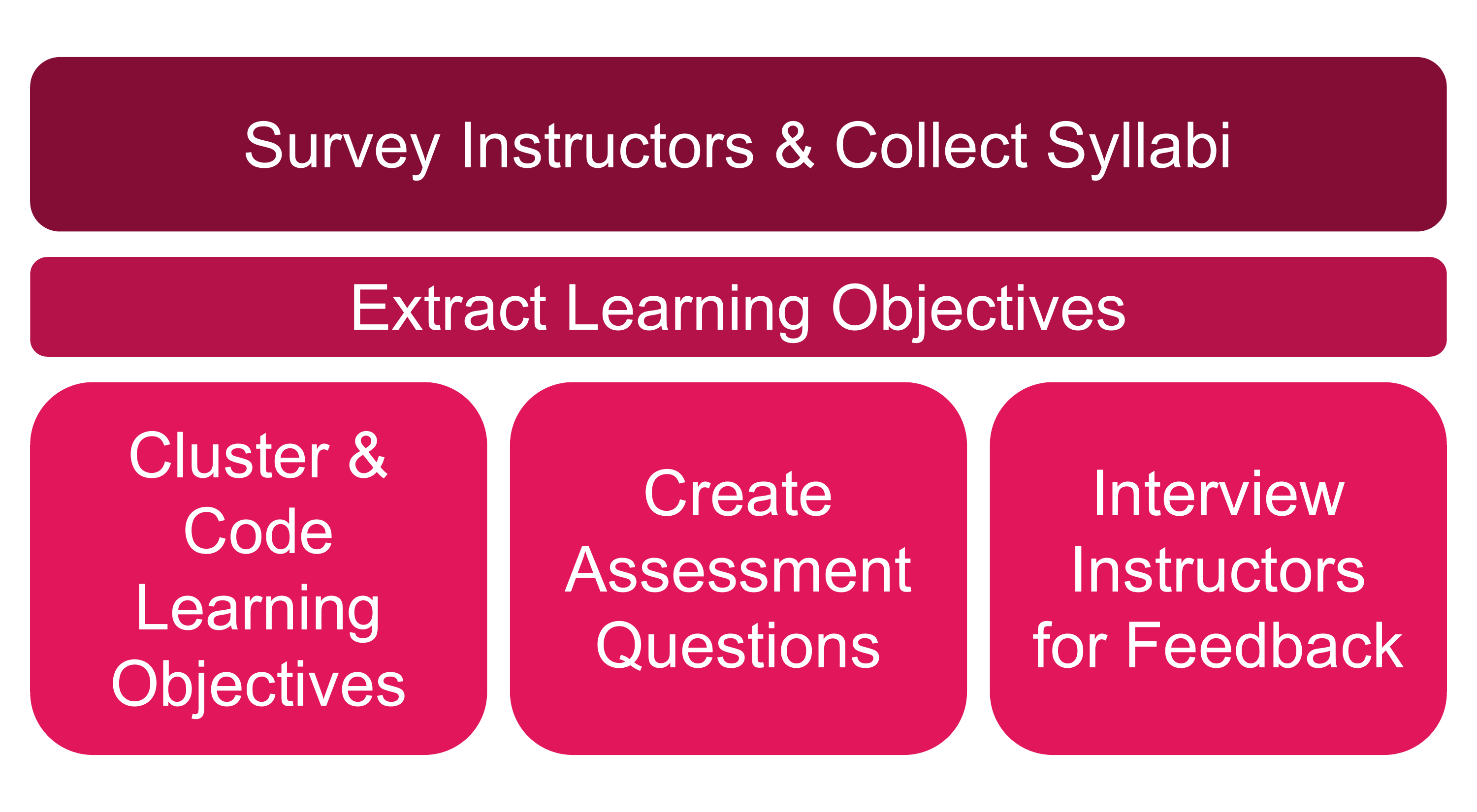}
    \caption{Our process to create the visualization design concept inventory.} 
    \label{fig:methodsoverview}
\end{figure}

We synthesize our concept inventory in two parts (see \autoref{fig:methodsoverview}). First, we collect syllabi from 38 visualization courses and 35 instructors. We extract the stated learning objectives from each syllabus, which we confirm with instructors.Then, we organize these learning objectives into a hierarchy representing the most important topics, skills, and relationships between them. Second, to articulate a more precise understanding of what each topic and skill represents, we develop example assessment questions that encapsulate them in a multiple choice format (i.e., the concept inventory). 
We evaluate our resulting topic/skills clusters and concept inventory questions through interviews with 10 visualization instructors. We followed an iterative approach, where we interleaved the interviews in parallel to our concept inventory development. In summary, this paper makes the following contributions:
\begin{itemize}[itemsep=0pt]
    \item We \textbf{present a concept inventory for visualization design}. To guide the concept inventory, we present a hierarchy of learning objectives that instructors prioritize when teaching visualization design. To form the concept inventory, we develop example questions to assess corresponding concepts and skills as well as common misconceptions in visualization design.
    \item We \textbf{introduce existing theory and methodology from the education research community} to a visualization audience and demonstrate its relevance to visualization research and education.
    \item We \textbf{present our synthesis of the concept inventory} through qualitative analysis of 38 course syllabi and interviews with 10 visualization course instructors. Our findings reveal \textbf{open questions and new opportunities} for visualization researchers. 
\end{itemize}

%% file: content/2-relatedwork.tex
\section{Background and Related Work}
\label{sec:related-work}

\subsection{Concept Inventories}
\label{sec:related-work:concept-inventories}

The idea of a concept inventory was first introduced for physics teachers to understand students' commonsense understanding of force through the Force Concept Inventory (FCI) \cite{forceCIHestenes}. In this work, the authors specifically recommended the use of a concept inventory as ``a diagnostic tool [\ldots] to identify and classify misconceptions'', and ``for evaluating instruction'' (p.~150). They also noted that a concept inventory has ``limited value'' as a placement examination, especially for beginning students.
Sands et al.~\cite{Sands2018CIsForUnderstanding} describe how concept inventories are different from final examinations and similar summative assessments and note that they are primarily about measuring understanding and not declarative knowledge. While the exact definition of concept inventories can vary~\cite{Epstein2013}, for our work, we follow two key characteristics of concept inventories---(1) as ``tests of the \textit{most basic} conceptual comprehension of foundations of a subject and not of computation skill,'' and (2) as ``different from final exams and mak[ing] no pretense of testing everything in a course.~\cite[p.~1019]{Epstein2013}'' In light of these two characteristics, though a typical visualization design course may include elements (e.g., domain-specific techniques) that are not covered by the concept inventory presented in this paper, we focus on the concepts that we identify as comprising the foundations of the subject. 

Since the introduction of FCI, similar multiple-choice tests have since been used to evaluate students' understanding of core topics and ideas in a wide variety of disciplines to understand students' abilities, to identify where students need support, and to evaluate the effectiveness of a course in addressing areas of initial weakness \cite{MurtazaCIReview, graphicalCommunicationCISteinhauer, Veith2022GroupTheoryCI, Salimpour2022CosmologyCI}. Within the field of computer science, concept inventories have been developed for introductory programming \cite{TewCSCI, caceffoCSCI, programmingCIreplication}, data structures \cite{Porter2019DataStructuresCI, DataStructuresCIBuildingDetails}, operating systems \cite{OSConceptInventory}, computer architecture \cite{ComputerArchCI}, discrete mathematics \cite{DiscreteMathCI}, and artificial intelligence \cite{Zhang2024DevelopingAICI}. In a paper that examines the use and impact of two concept inventories (FCS1 and SCS1) in computer science, Parker et al.~\cite{10.1145/3446871.3469744} describe a range of uses such as measuring performance differences between multiple courses~\cite{tew2010assessing}, replicating prior work~\cite{10.1145/2543882.2543884}, comparing teaching strategies, understanding self-directed learning with an online tool~\cite{10.1145/3386527.3405928}, and developing additional measures \cite{10.1145/3291279.3339407}. Beyond computer science, uses of concept inventories include comparison of different curricula for mechanics courses \cite{10.1119/1.3703517} and student assessment and faculty professional development in the life sciences~\cite{doi:10.1187/cbe.10-05-0069}.

Although concept inventories are by definition assessments, they are typically developed first as theories, essentially the researchers' hypothesis of what skills, concepts, and misconceptions should be recorded for a given topic and discipline \cite{Adams01062011, 10.1145/1352135.1352226, MurtazaCIReview, GoldmanSettingScopeOfCIs}. Then, follow-up projects are conducted to evaluate the validity and generalizability of the theory through multiple classroom deployments and other frameworks\cite{Adams01062011, 10.1145/3287324.3287370, programmingCIreplication, Jorion2015-CIValidationFramework, MurtazaCIReview}. In this paper, we focus on the first part of the concept inventory development process, introducing a theory of what skills and concepts instructors focus on when teaching visualization design.

The concept inventory itself is typically a multi-question assessment, where each question is multiple choice rather than free response~\cite{Sands2018CIsForUnderstanding, Furrow2019}. While multiple choice questionnaires are easier to administer and assess, in a 2018 paper, Sands et al.~note ways in which they also fall short, and describe  in-progress work in this area that aims to move beyond such a format (e.g., by requiring students to provide free-text responses)~\cite{Sands2018CIsForUnderstanding}. Given that our research goal is primarily about a concept inventory for visualization design, and not about advancements in the science of concept inventories, we chose to adhere to the more common multi-question assessment format. Furthermore, the fixed-answer questionnaire format aligns with existing assessments in visualization literacy research, including multiple-choice questions on chart interpretation (e.g., Mini-VLAT~\cite{minivlat},  VLAT~\cite{vlat}, CALVI~\cite{ge2023calvi}) and true/false questions on chart construction (e.g., AVEC~\cite{ge2025avec}.
Each question in a concept inventory typically has three key parts \cite{MurtazaCIReview}:
\begin{enumerate}[itemsep=-2pt]
    \item \textbf{The Question:} these should cover a core concept for the topic the concept inventory has been developed for.
    \item \textbf{A Correct Answer:} there should be only a single correct answer for each question.
    \item \textbf{Distractors}: these are incorrect answers corresponding to common conceptual mistakes related to the topic. 
\end{enumerate}

Our work follows the high-level process utilized by Tew and Guzdial in developing the Foundational CS1 (FCS1) Assessment~\cite{TewCSCI, 6562691} for introductory Computer Science education. In this process, Tew and Guzdial initially published a method to develop a language-independent assessment for  CS1-level concepts by using CS1 textbooks as a starting point and then engaging with a panel of experts in CS education to further refine the test specification~\cite{10.1145/1734263.1734297}. This publication also included a report of the authors' progress till-date in following the methods that they described. They then followed up with an empirical study~\cite{TewCSCI} to validate the assessment that was ultimately developed using the methods described in the earlier paper. A similar two-stage process was also followed to develop a concept inventory for Digital Logic~\cite{10.1145/1352322.1352226, 10.1145/1734263.1734298}. 

\subsection{Visualization Education and Assessment}

Over the years, much research has been done on what students need from visualization courses in order to maximize its benefits \cite{Camm2022WhatToTeachInVis, Bandi2017CraftingVisCourseForIndustry, Pucha2016CriticalThinkingActivitiesVisCourse}. This has included the benefits of embedding real-world problem solving in course structure \cite{Nestorov2019VisCourseWithRealApp}, how to best teach students in particular fields \cite{Jnicke2019VisForJournalismStudents, Hudiburgh2020VisForStatsStudents}, identifying common mistakes and pitfalls to suggest changes in visualization education \cite{visPitfallsinSciPubs, fox2020surfacingmisconceptionsvisualizationcritique}, preparing students for the professional workforce~\cite{ryan2019teaching}, and how visualization education needs to respond to major technological changes such as AI \cite{Bach2023VisEdChallenges}. Other research has studied how visualization classes improve visualization literacy skills \cite{whatStudentsLearnInVisHedayati}.

Evaluation tools for visualization education often test a person's understanding of a single aspect of visualization. For example, visualization literacy assessments such as VLAT~\cite{vlat} and Mini-VLAT~\cite{minivlat} test a person's ability to interpret different chart types by asking the person to answer multiple-choice data analysis questions using corresponding charts. 
CALVI measures how well people can interpret misleading visualization designs~\cite{ge2023calvi}.
Inspired by existing frameworks for visualization literacy~\cite{borner2019data}, AVEC~\cite{ge2025avec} assesses a person's ability to construct an effective chart to complete one of three common visualization analysis tasks: describing distributions, finding trends and correlations, and comparing patterns/distributions within a dataset. Visualization literacy assessments have also been used to measure what students have learned from data visualization courses, but show limited benefits thus far~\cite{whatStudentsLearnInVisHedayati}.
As an alternative approach, VisGrader allows instructors to define unit tests with which to automatically grade D3-based assignment submissions and provide students with feedback~\cite{hull2023visgraderautomaticgradingd3}.

We acknowledge that alternative assessment strategies in visualization could also be applied to visualization education, such as those in research on visualization linting (e.g., \cite{mcnutt2018Linting, 2025-color-buddy, hopkins2020visualint}), graphical perception (e.g., \cite{zeng2023review,quadri2022visualization,kim2026data}), chart comprehension (e.g., \cite{quadri2024see}), and data exploration (e.g., \cite{feng2019patterns,wootton2025charting,battle2019characterizing}) as well as visualization research pertaining to learning objectives (e.g., \cite{lee-robbins2022learning,lee-robbins2023affective,adar2021communicative,burns2020how}). Our concept inventory questions are inspired in part by existing assessment methods outside of visualization education. However, our focus in this paper is on complementing existing education and literacy-focused assessments.

We argue that two factors distinguish a potential concept inventory for visualization design from existing visualization assessments. First, concept inventory questions strategically incorporate distractor answers reflecting common misconceptions within the given topic, situating them alongside positive answers representing best practices. To the best of our knowledge, existing assessments do not cover common misconceptions in visualization design (although they do cover interpretation misconceptions, e.g., CALVI~\cite{ge2023calvi}). Second, concept inventories are grounded in the teaching practices of instructors and what they aim to convey to students through a course, making concept inventories broader than existing visualization literacy/education assessments.

\subsection{Positionality Statement}

 Of the visualization instructors surveyed and interviewed, some have personal connections to members of the research team. These relationships include colleagues, former and current advisors and advisees, and collaborators. We recognize that these factors could introduce some bias. Additionally, with respect to coding the course syllabi, the coders already had some themes in mind due to their prior understanding of the different skills that are developed in visualization courses. That being said, the research team took multiple precautions to mitigate potential bias in our analysis process, such as soliciting syllabi from a large fraction of the visualization community (114 instructors were invited to participate in our study), considering the geographic diversity of instructors invited (instructors from five continents were invited to participate), inviting all consenting instructors for an interview as part of our study, and opting to cluster all observed learning objectives from all participating instructors through our analysis process.

%% file: content/3-methods.tex
\section{Methods Overview}
\label{methods}

In this section, we summarize our design goals and provide a brief overview of our research methods (see  \autoref{fig:methodsoverview}).
The survey and interview process were reviewed by University of Washington's IRB and determined to be exempt.

\subsection{Design Goals}
\label{sec:methods:design-goals}

We summarize three design goals that drive our methods:

\paragraph{\textbf{DG1}: Center Instructors’ Teaching Goals.} We observe that current assessments do not seem to emphasize feedback from a range of visualization course instructors. We believe this limits the relevance of current assessments to visualization courses. For example, recent work shows how the VLAT may not reflect what students learn in a visualization course and instead may reflect prerequisite knowledge~\cite{whatStudentsLearnInVisHedayati}. Thus, a goal of our concept inventory is to emphasize the topics and skills that instructors want students to learn through their courses.

\paragraph{\textbf{DG2}: Emphasize Skill Breadth over Depth.} While targeted assessments are essential for measuring specific skills (e.g., detecting misinformation~\cite{ge2023calvi}, constructing a visualization~\cite{ge2025avec}), we observe a lack of focus on the \emph{breadth} of skills taught in visualization courses. Revisiting our first research question, what are skills that instructors believe are necessary to produce effective visualization designers? Current work is unable to provide a comprehensive answer. In response, we aim to discover the ``convex hull’’ of visualization design skills that instructors aim to teach students through their courses.

\paragraph{\textbf{DG3}: Emphasize Strategic Examples Over Full Assessment Coverage.} To answer our second research question, we need to go beyond topic/skills clusters to develop assessment questions for visualization design. That being said, we acknowledge that creating an in-depth assessment covering all observed skills is impractical.
It could take hours for students to complete such a comprehensive assessment, given the time required to assess individual skills using VLAT (15 minutes~\cite{vlat}), AVEC (22 minutes~\cite{ge2025avec}), and CALVI (20 minutes~\cite{ge2023calvi}). Thus, maximizing for breadth (i.e., achieving \dgtwo) necessitates a tradeoff in the depth of knowledge that individual assessment questions can cover for any one topic. Rather than aiming for complete and in-depth coverage of all observed skills, we instead treat the assessment questions in the visualization design concept inventory as cornerstone examples for assessing key skills within each cluster/subcluster, in line with prior work (see \autoref{sec:related-work:concept-inventories} for details). As a result, the concept inventory complements existing depth-oriented assessments with a breadth-oriented assessment covering multiple visualization design skills taught in visualization courses. Furthermore, the concept inventory can guide the development of new, in-depth visualization design assessments in future work, which we discuss in \autoref{sec:discussion:future-research}. 

\subsection{Surveying Visualization Instructors}
\label{surveymethods}

To achieve \dgone, we surveyed visualization instructors to get insight into their intent behind developing their syllabi. Potential participants were identified through the professional networks of the authors and reviewing CSRankings\footnote{https://csrankings.org/} for visualization research. 
Eligibility was confirmed by checking for visualization courses taught in recent years. Instructors were invited over email to participate in the survey.

\begin{table}[]
\centering
\caption{The final codes from the qualitative coding of the learning objectives. Rare skills and themes were not coded.}
\label{tab:syllabuscodes}
\begin{tabular}{p{0.075\linewidth} | p{0.8\linewidth}}
\hline
\textbf{Code} & \textbf{Description}   \\ \hline
VD  & Understanding or applying visualization design \\
\rowcolor[HTML]{EFEFEF} 
PA  &  Programming and application where the development of technical skills is emphasized \\
DL  &  Data literacy \\
\rowcolor[HTML]{EFEFEF} 
CE  & Critiquing and evalutating visualizations \\
VR  & Visualization research - Reading, discussing, and critiquing visualization research \\
\rowcolor[HTML]{EFEFEF} 
CS  & Communication skills - Developing writing and presentation skills, group work \\ 
WF  & Project workflow and design process \\
\rowcolor[HTML]{EFEFEF} 
RW  & Real-world problem solving, including working with real-world data and data storytelling \\ 
SD  & Designing and developing for specific data domains, including vector fields and 3D modeling \\
\rowcolor[HTML]{EFEFEF} 
VL  & Visualization Literacy \\
DA  & Data analysis and wrangling \\\hline
\end{tabular}
\end{table}

In total, \textbf{we invited 114 instructors to participate}, 27 from private and 87 from public universities. We invited 97 instructors from North America (90 - US, 7 - Canada) and 27 outside of North America (16 - Europe, 6 - Asia, 4 - Oceania, 1 - South America).
Of those invited, \textbf{35 instructors consented to the inclusion of their syllabi in our analysis}, with 3 instructors providing syllabi for two courses each. All respondents are located in the USA or Canada. 27 instructors are based in computer science departments,
5 in information science departments, and 3 instructors in other departments. Participants exhibit a wide range of teaching experience from teaching visualization for the first time to 30 years of experience teaching visualization. All respondents have a masters or doctoral terminal degree.

Respondents were surveyed regarding their educational and teaching background, asked to provide their most recent visualization course syllabus, and asked specific questions regarding the development of that syllabus. Of the most important questions were
\textit{"How many years have you been teaching visualization courses?"} and 
\textit{"What are the most important learning outcomes for students enrolled in your course?"}

In total, \textbf{38 syllabi were analyzed in our study}, where 7 syllabi were submitted via email and 31 were submitted via the survey. The syllabi include 17 undergraduate courses, 10 graduate courses, and 11 courses including both undergraduate and graduate students. All submitted syllabi were from courses offered within the past two years (Spring 2024 to Spring 2026), except for one course offered in 2021. Course length varied with one course scheduled for 8 weeks, five courses scheduled for 10 weeks, five courses scheduled for 13 weeks, and 27 courses scheduled for 15 weeks.
We only analyze syllabi from instructors who consented to participating in our study as course syllabi are considered the intellectual property of the course creator. As the participation of some visualization instructors cannot be fully anonymized due to their online teaching presence, we gave our participants the choice to be credited by name (as recommended by Bruckman et al. \cite{usingrealnames}) and to have their syllabi included in the supplemental materials.

\subsection{Coding and Clustering Learning Objectives}

Clustering learning objectives by common themes enables us to identify the skills and concepts introduced through each course, filter for those that are specific to visualization design, and assess their prevalence across multiple courses, enabling us to achieve \dgtwo.

\paragraph{Extracting and Validating Course Objectives.} To extract learning objectives from course syllabi, we selected bullet points detailing course goals. These bullet points were often listed under sections with titles such as \textit{"Learning Objectives"} or \textit{"Course Outcomes"}, or these bullet points appeared after phrases such as \textit{“By the end of this course, students will be able to...”}. To ensure we did not miss any learning objectives, we compared instructors' survey responses regarding current/additional learning objectives to the corresponding syllabus. Nearly all survey responses either briefly summarized the learning objectives from their syllabi (10 out of 31 respondents), directly copy-pasted from their syllabi (8/31), did not provide a response to this question (5/31), or simply referred us to their syllabus (5/31). Only 3 respondents provided different learning objectives from their syllabus, which we added to our analysis.
The significant overlap between survey responses and course syllabi confirmed that the learning objectives listed in the syllabi were reflective of what instructors expected students to learn from the course. Therefore, analyzing the learning objectives from the syllabi should be sufficient to identify and understand the core skills and topics covered in visualization courses.

\paragraph{Coding and Clustering Learning Objectives.} 
All extracted learning objectives were clustered according to common themes, goals, and topics. 
The clusters and subclusters were iteratively refined and reorganized through team discussions and continuously updated using feedback from interviews with instructors.
Initially, we clustered the learning objectives according to broad themes, but as more learning objectives were added, some clusters were refined into smaller subclusters.
Instructors often wrote compound learning objectives, covering multiple themes within a single objective. In these cases, care was taken to tease apart distinct themes.
Consider the following example: \emph{``Design and construct simple static visualizations.''} This learning objective would be labeled with the themes VD (understanding or applying visualization design) and PA (programming and application) from \autoref{tab:syllabuscodes}. 
To preserve the multifaceted nature of compound learning objectives in our clustering, we duplicated learning objectives in multiple clusters. Continuing our example above, this learning objective would fall under two clusters: Designing Visualizations and Developing Visualizations.

\begin{table}[]
\centering
\caption{University, role, teaching experience, and course information for interview participants. Some participants discussed two courses taught.}
\label{tab:interview-participants}
\begin{tabular}{lllll}
\hline
\textbf{PID} & \textbf{Role}  & \textbf{\begin{tabular}[c]{@{}l@{}l@{}}Years\\ Teaching \\ Vis\end{tabular}} & \textbf{\begin{tabular}[c]{@{}l@{}l@{}}Public or\\Private\\ University\end{tabular}} & \textbf{Ugrad/Grad} \\ \hline
P1    & Assist. Prof.  & 1 & Public     & Ugrad               \\
\rowcolor[HTML]{EFEFEF} 
P2    & Assist. Prof.  & 5 & Public    & Ugrad/Grad          \\
P3    & Assist. Prof.  & 6 & Public    & Ugrad, Grad         \\
\rowcolor[HTML]{EFEFEF} 
P4    & Assist. Prof.  & 1 & Public    & Ugrad/Grad         \\
P5    & Assoc. Prof.   & 14  & Private   & Grad                \\
\rowcolor[HTML]{EFEFEF} 
P6    & Assoc. Prof.   & 12 & Public    & Ugrad               \\
P7    & Assoc. Prof.   & 12 & Public    & Ugrad, Ugrad        \\
\rowcolor[HTML]{EFEFEF} 
P8    & Professor      & 25 & Public    & Grad                \\ 
P9    & Professor      & 24 & Public    & Ugrad               \\ 
\rowcolor[HTML]{EFEFEF} 
P10   & Assist. Prof.  & 1 & Public    & Ugrad/Grad           \\ \hline
\end{tabular}
\end{table}

\begin{table*}[t]
\caption{Overview of how the learning objectives were clustered, showing top-level clusters, subclusters within the top-level clusters, our interpretation of each subcluster (i.e., what students should be able to do), an example of what types of learning objectives could be included in each subcluster, and the cognitive skills engaged by each subcluster. Clusters in the grey rows are excluded from the concept inventory.}
\label{tab:clusterstable}
\resizebox{\textwidth}{!}{%
\begin{tabular}{llll}
\hline
\rowcolor[HTML]{FFFFFF} 
\textbf{Cluster} & \textbf{Subcluster} & \textbf{Subcluster Interpretation and Example Learning Objective} & \textbf{Cognitive Skill Engaged} \\ \hline
\rowcolor[HTML]{FFFFFF} 
\cellcolor[HTML]{FFFFFF} & Practicing Evaluation & \begin{tabular}[c]{@{}l@{}}\textbf{Interpretation:} Evaluate/Critique a visualization according to best practices, effectiveness,\\ and suitability for the audience.\\ \textbf{Example:} Critically evaluate the visualization design for effectiveness and \\ interpretability, taking the intended audience into account.\end{tabular} & \begin{tabular}[c]{@{}l@{}}understand, analyze, \\ evaluate\end{tabular} \\ \cline{2-4} 
\rowcolor[HTML]{FFFFFF} 
\cellcolor[HTML]{FFFFFF} & Theory-driven Evaluation & \begin{tabular}[c]{@{}l@{}}\textbf{Interpretation:} Evaluate/Critique a visualization through a theoretical lens.\\ \textbf{Example:} Critically evaluate visualizations using ideas from visualization theory, \\ quantitative methods, and qualitative methods.\end{tabular} & \begin{tabular}[c]{@{}l@{}}understand, analyze, \\ evaluate\end{tabular} \\ \cline{2-4} 
\rowcolor[HTML]{FFFFFF} 
\multirow{-3}{*}{\cellcolor[HTML]{FFFFFF}\begin{tabular}[c]{@{}l@{}}Critiquing \& Evaluating \\ Design\end{tabular}} & \begin{tabular}[c]{@{}l@{}}Redesigning \\ Visualizations\end{tabular} & \begin{tabular}[c]{@{}l@{}}\textbf{Interpretation:} Correct identified design issues or improve the design according to \\ specified goals and the target audience.\\ \textbf{Example:} Suggest improvements/alternative design solutions to visualization flaws.\end{tabular} & \begin{tabular}[c]{@{}l@{}}understand, analyze, \\ evaluate, apply\end{tabular} \\ \hline
\rowcolor[HTML]{FFFFFF} 
\cellcolor[HTML]{FFFFFF} & Overview of Field & \begin{tabular}[c]{@{}l@{}}\textbf{Interpretation:} Understand how the field of visualization draws on other fields.\\ \textbf{Example:} Become familiar with the visualization field and its subdisciplines.\end{tabular} & remember, understand \\ \cline{2-4} 
\rowcolor[HTML]{FFFFFF} 
\cellcolor[HTML]{FFFFFF} & Understanding Theory & \begin{tabular}[c]{@{}l@{}}\textbf{Interpretation:} Understand the core and foundational visualization theories.\\ \textbf{Example:} Develop an understanding of core visualization theory, including\\ human perception and cognition.\end{tabular} & remember, understand \\ \cline{2-4} 
\rowcolor[HTML]{FFFFFF} 
\multirow{-3}{*}{\cellcolor[HTML]{FFFFFF}\begin{tabular}[c]{@{}l@{}}Understanding \\ Visualization \\ Design Principles\end{tabular}} & Understanding Methods & \begin{tabular}[c]{@{}l@{}}\textbf{Interpretation:} Understand key techniques, best practices, and how to apply vis. theory. \\ \textbf{Example:} Develop an understanding of key visualization techniques and \\ best practices.\end{tabular} & remember, understand \\ \hline
\rowcolor[HTML]{FFFFFF} 
\cellcolor[HTML]{FFFFFF} & \begin{tabular}[c]{@{}l@{}}Designing Effective \\ Visualizations\end{tabular} & \begin{tabular}[c]{@{}l@{}}\textbf{Interpretation:} Implement effective design choices to achieve visualization goals.\\ \textbf{Example:} Design visualizations that utilize effective design to communicate data insights.\end{tabular} & \begin{tabular}[c]{@{}l@{}}understand, analyze, \\ apply, create\end{tabular} \\ \cline{2-4} 
\rowcolor[HTML]{FFFFFF} 
\cellcolor[HTML]{FFFFFF} & \begin{tabular}[c]{@{}l@{}}Designing Original \\ Visualizations\end{tabular} & \begin{tabular}[c]{@{}l@{}}\textbf{Interpretation:} Design basic, simple visualizations (excludes bespoke visualizations).\\ \textbf{Example:} Ideate and sketch novel/original visualizations.\end{tabular} & \begin{tabular}[c]{@{}l@{}}understand, apply, \\ create\end{tabular} \\ \cline{2-4} 
\rowcolor[HTML]{FFFFFF} 
\cellcolor[HTML]{FFFFFF} & \begin{tabular}[c]{@{}l@{}}Designing for \\ Communication\end{tabular} & \begin{tabular}[c]{@{}l@{}}\textbf{Interpretation:} Design visualizations to answer a question/communicate a specific insight.\\ \textbf{Example:} Design visualizations for data storytelling from data.\end{tabular} & \begin{tabular}[c]{@{}l@{}}understand, analyze,\\ apply, create\end{tabular} \\ \cline{2-4} 
\rowcolor[HTML]{FFFFFF} 
\multirow{-4}{*}{\cellcolor[HTML]{FFFFFF}Designing Visualizations} & \begin{tabular}[c]{@{}l@{}}Understanding the\\ Design Process\end{tabular} & \begin{tabular}[c]{@{}l@{}}\textbf{Interpretation:} Understand a typical workflow to complete a visualization project.\\ \textbf{Example:} Understand, articulate, and apply a structured design process.\end{tabular} & \begin{tabular}[c]{@{}l@{}}remember, understand,\\ apply\end{tabular} \\ \hline
\rowcolor[HTML]{EFEFEF} 
\cellcolor[HTML]{EFEFEF} & \begin{tabular}[c]{@{}l@{}}Designing/Implementing \\ Interactivity\end{tabular} & \begin{tabular}[c]{@{}l@{}}\textbf{Interpretation:} Design and create an interactive visualization.\\ \textbf{Example:} Design and implement interactive visualizations using D3.js.\end{tabular} & \begin{tabular}[c]{@{}l@{}}understand, apply,\\ create\end{tabular} \\ \cline{2-4} 
\rowcolor[HTML]{EFEFEF} 
\multirow{-2}{*}{\cellcolor[HTML]{EFEFEF}\begin{tabular}[c]{@{}l@{}}Designing Visualizations /\\ Developing Visualizations\end{tabular}} & \begin{tabular}[c]{@{}l@{}}Designing/Building for \\ Specific Data Domains\end{tabular} & \begin{tabular}[c]{@{}l@{}}\textbf{Interpretation:} Understand the unique challenges and techniques associated with \\ specific data domains and design/create visualizations accordingly.\\ \textbf{Example:} Design and implement visualizations for 3D data and scalar fields.\end{tabular} & \begin{tabular}[c]{@{}l@{}}understand, apply, \\ create\end{tabular} \\ \hline
\rowcolor[HTML]{EFEFEF} 
Developing Visualizations & Authoring Visualizations & \begin{tabular}[c]{@{}l@{}}\textbf{Interpretation:} Use a visualization tool to create/program a visualization.\\ \textbf{Example:} Implement web-based visualizations with multiple tools.\end{tabular} & \begin{tabular}[c]{@{}l@{}}understand, apply, \\ create\end{tabular} \\ \hline
\rowcolor[HTML]{EFEFEF} 
\begin{tabular}[c]{@{}l@{}}Data Transformation \& \\ Analysis\end{tabular} & --- & \begin{tabular}[c]{@{}l@{}}\textbf{Interpretation:} Clean and manipulate the data for a visualization.\\ \textbf{Example:} Utilize data science techniques and tools to clean, filter, and manipulate data.\end{tabular} & understand, apply \\ \hline
\rowcolor[HTML]{EFEFEF} 
\cellcolor[HTML]{EFEFEF} & Ethics & \begin{tabular}[c]{@{}l@{}}\textbf{Interpretation:} Read about and discuss ethical issues in the field of visualization.\\ \textbf{Example:} Explain the ethical foundations of core principles in visualization.\end{tabular} & \begin{tabular}[c]{@{}l@{}}remember, understand,\\ analyze\end{tabular} \\ \cline{2-4} 
\rowcolor[HTML]{EFEFEF} 
\multirow{-3}{*}{\cellcolor[HTML]{EFEFEF}\begin{tabular}[c]{@{}l@{}}Understanding \\ Visualization \\ Research\end{tabular}} & \begin{tabular}[c]{@{}l@{}}General Visualization\\ Research\end{tabular} & \begin{tabular}[c]{@{}l@{}}\textbf{Interpretation:} Read and discuss visualization research papers.\\ \textbf{Example:} Read, discuss, and engage with fundamental visualization research.\end{tabular} & \begin{tabular}[c]{@{}l@{}}remember, understand,\\ analyze\end{tabular} \\ \hline
\end{tabular}%
}
\end{table*}

\paragraph{Coding Consistency/Reliability.} Two members of the research team independently coded learning objectives from a (separate) third of the syllabi, after which the entire author team met to discuss the codebook. Then the two members independently coded the remaining third of the syllabi, achieving a Cohen’s Kappa inter-rater reliability score of 0.92 \cite{cohen1960coefficient}. Remaining discrepancies were manually resolved through discussion.  The final theme labels are shown in \autoref{tab:syllabuscodes} and the final clusters are shown in  \autoref{tab:clusterstable}. We present our results in \autoref{analysis}.

\subsection{Developing Concept Inventory Questions}

To achieve \dgthree, we developed example assessment questions for each skill cluster commonly associated with visualization design. 
To respect the intellectual property of our survey respondents, we did not rely on their course materials to develop the concept inventory questions. Instead, we draw on established concepts and best practices found in published visualization research papers and textbooks to inform the design of our assessment questions. We detail our question design strategy and rationale in \autoref{vdconceptinventory}.

\paragraph{Question Evaluation and Refinement.} We invited our 35 survey participants to provide feedback on our topic clusters and concept inventory; 10 agreed to be interviewed (see \autoref{tab:interview-participants}). The audio and video for these interviews were recorded with the consent of the interviewees.
We aimed to solicit feedback on the concept inventory as early as possible, enabling instructors to evaluate our topic/skills clusters and example assessment questions and provide feedback throughout our synthesis process.
In between interviews, we used the feedback we received to improve the concept inventory, i.e., to add missing questions, refine existing questions, and remove irrelevant questions.

Interviewees were asked to review the clusters of learning objectives with their own learning objectives highlighted to contextualize the data. Interviewees were asked about cluster quality and which clusters should be included in the concept inventory. Then, interviewees were asked to evaluate the concept inventory; specifically, if there was anything misleading or confusing about the question wording and answer options, the relevance of such questions in evaluating the corresponding cluster of learning objectives, and what they would change or do differently.

\paragraph{Feedback Saturation.} While many questions could be developed to assess visualization design beyond what we propose in our concept inventory, we observe that a form of qualitative data saturation could be inferred if feedback from instructors shifted from substantial, i.e., requesting major changes in question design and the addition of entire questions, to subtle, i.e., requesting small and incremental refinements to existing questions and adding alternative versions to support specific student audiences, e.g., color blind and dyslexic students. 
By the end of our interview process, we were no longer making significant changes to the questions and instead shifting towards smaller, incremental improvements, i.e., clarifying terminology and ensuring the context of each question and answer was clear.
Thus, we argue that our design process came close to reaching a point of saturation when comparing the feedback from our earliest and latest interviews. 

%% file: content/4-analysis.tex
\section{Analysis of Clusters}
\label{analysis}

\autoref{tab:clusterstable} summarizes the key skills and concepts that instructors emphasize in their visualization courses, answering our first research question. The final skill/task hierarchy includes seven top-level clusters and 15 subclusters.
To ensure clarity regarding what the subclusters cover, we include our interpretation of each in \autoref{tab:clusterstable}.

\begin{table*}[]
\centering
\caption{Mapping of concept inventory questions to corresponding topic and skills subclusters.}
\label{tab:subcluster-question-mapping}
\begin{tabular}{lllllllllllllll}
\hline
                                   & \textbf{1}    & \textbf{2}    & \textbf{3}    & \textbf{4}    & \textbf{5}    & \textbf{6}    & \textbf{7}    & \textbf{8}    & \textbf{9}    & \textbf{10}   & \textbf{11}   & \textbf{12}   & \textbf{13}   & \textbf{14}   \\ \hline
\textbf{[Critique \& Eval]} Practicing Evaluation              & \checkmark & \checkmark &      &      &      &      &      &      &      &      &      &      &      &      \\
\rowcolor[HTML]{EFEFEF} 
\textbf{[Critique \& Eval]} Theory-driven Evaluation           &      &      & \checkmark &      &      &      &      &      &      &      &      &      &      &      \\
\textbf{[Critique \& Eval]} Redesigning Visualizations         & \checkmark &      &      & \checkmark &      & \checkmark &      &      &      &      &      &      &      &      \\
\rowcolor[HTML]{EFEFEF} 
\textbf{[Understand Principes]} Overview of Field                  &      &      &      &      &      &      &      &      &      & \checkmark & \checkmark &      &      &      \\
\textbf{[Understand Principes]} Understanding Theory               &      &      &      &      &      &      &      &      &      &      &      & \checkmark & \checkmark & \checkmark \\
\rowcolor[HTML]{EFEFEF} 
\textbf{[Understand Principes]} Understanding Methods              &      &      &      &      &      &      &      &      &      &      &      &      &      & \checkmark \\
\textbf{[Design Vis]} Designing Effective Visualizations &      &      &      &      & \checkmark & \checkmark & \checkmark &      & \checkmark &      &      &      &      &  \checkmark    \\
\rowcolor[HTML]{EFEFEF} 
\textbf{[Design Vis]} Designing Original Visualizations  &      &      &      &      & \checkmark & \checkmark     &      &      &      &      &      &      &      &      \\
\textbf{[Design Vis]} Designing for Communication         &      &      &      &      &      &      &      &      & \checkmark &      &      &      &      &      \\
\rowcolor[HTML]{EFEFEF} 
\textbf{[Design Vis]} Understanding the Design Process   &      &      &      &      &      &      &      & \checkmark &      &      &      &      &      &      \\ \hline
\end{tabular}
\end{table*}

\subsection{Clusters/Subclusters Included in the Concept Inventory}
\label{sec:clusters:includedInCI}

Of the seven top-level clusters, we focus on the three most relevant to visualization design according to our qualitative codes and instructor feedback: \textbf{Critiquing \& Evaluating Design}, \textbf{Understanding Visualization Design Principles}, and  \textbf{Designing Visualizations}.
Initially, \textbf{Developing Visualizations} was also included in our concept inventory.
However, we excluded this cluster based on feedback from our interviews as instructors considered programming skills to be orthogonal to visualization design.
Similarly, while the \textbf{Designing/Implementing Interactivity} and \textbf{Designing/Building for Specific Data Domains} are subclusters of both the \textbf{Designing Visualizations} and \textbf{Developing Visualizations} clusters, they were also considered more specialized skills outside the purview of the concept inventory.
\textbf{Data Transformation \& Analysis} was not considered a focal point in visualization design, as it overlaps heavily with other courses in data science and data management and varies widely in the tools and methods used to manipulate the data. \textbf{Understanding Visualization Research} was also considered out of scope, considering how many visualization research methods go far beyond the basic skills of visualization design.

\subsection{Cognitive Skills Engaged by Each Subcluster}

To gain a deeper understanding of which cognitive skills might be engaged within each subcluster, 
we qualitatively coded them using Bloom's taxonomy~\cite{Bloom_1956}, drawing on relevant work in visualization for further contextualization~\cite{designInfoVisCourse,burns2020how} (see  \autoref{tab:clusterstable}).
Note that while “Remember” is typically engaged across all subclusters, we only include it when it seems to be emphasized by a particular subcluster.
For the 10 subclusters focusing on visualization design, understanding visualization design is a core skill for all of them. Subclusters within the same cluster tend to engage the same cognitive skills with the \textbf{Designing Visualizations} subclusters relying heavily on applying visualization design skills and creating visualizations, the \textbf{Understanding Visualization Design Principles} subclusters focusing on remembering and understanding design principles, and the \textbf{Critiquing \& Evaluating Design} subclusters focusing on analyzing and evaluating visualizations. Overall, these findings align with what we would expect from the top-level clusters. In addition to these shared cognitive skills for each cluster, various subclusters may express slight differences in cognitive skills due to the specific learning objectives within each subcluster.

%% file: content/5-conceptinventory.tex
\section{Concept Inventory Design \& Rationale}
\label{vdconceptinventory}

To address our second research question, we developed a 14 question concept inventory.
\autoref{tab:subcluster-question-mapping} shows how each question maps to corresponding subclusters.
In this section, we describe our design rationale, organized around each subcluster listed in \autoref{tab:subcluster-question-mapping}. We discuss one example question per subcluster and only show a few illustrations across all subclusters to save space. Note that some subclusters draw on overlapping skills. Thus, some questions appear in multiple subclusters.

\subsection{Critiquing \& Evaluating Design}
\label{sec:concept-inventory:critiquing-evaluating-design}

Drawing on our observations from \autoref{analysis}, we find that the learning objectives in this cluster tend to emphasize similar levels of Bloom's taxonomy, which we operationalize in assessment questions as follows: \emph{understanding} common visualization design strategies or established theoretical principles, \emph{analyzing} the design decisions made within a current visualization and suggested alternatives, and \emph{evaluating} these design choices by ranking them according to specified design goals. Covering only the lower levels of understanding is insufficient to answer each question due to our selected answer distractors. For example, the ability to isolate design decisions within in a visualization is not enough. Students must also be able to rank these design choices to identify the design that best aligns with the specified design goals.

\begin{figure*}[t]
  \begin{subfigure}{0.325\textwidth}
    \includegraphics[alt={A breakdown of answer options for the question in Q1 (Figure 1). Option A: The y-axis should not be truncated. This is the correct  answer. This is a common deceptive design trick. While truncated axes can be appropriate, this should be clearly indicated at least. Option B: Color coding is unnecessary. This is a distractor. Valid critique but it doesn’t impact a viewer’s ability to accurately compare the GDP of the countries. Option C: Color legend is missing. This is a distractor. Valid critique but it doesn’t impact a viewer’s ability to accurately compare the GDP of the countries. Option D: GDP needs a currency.  This is a distractor. Valid critique as there is no unit or currency listed, but it doesn’t impact a viewer’s ability to accurately compare the GDP of the countries.}, width=\linewidth]{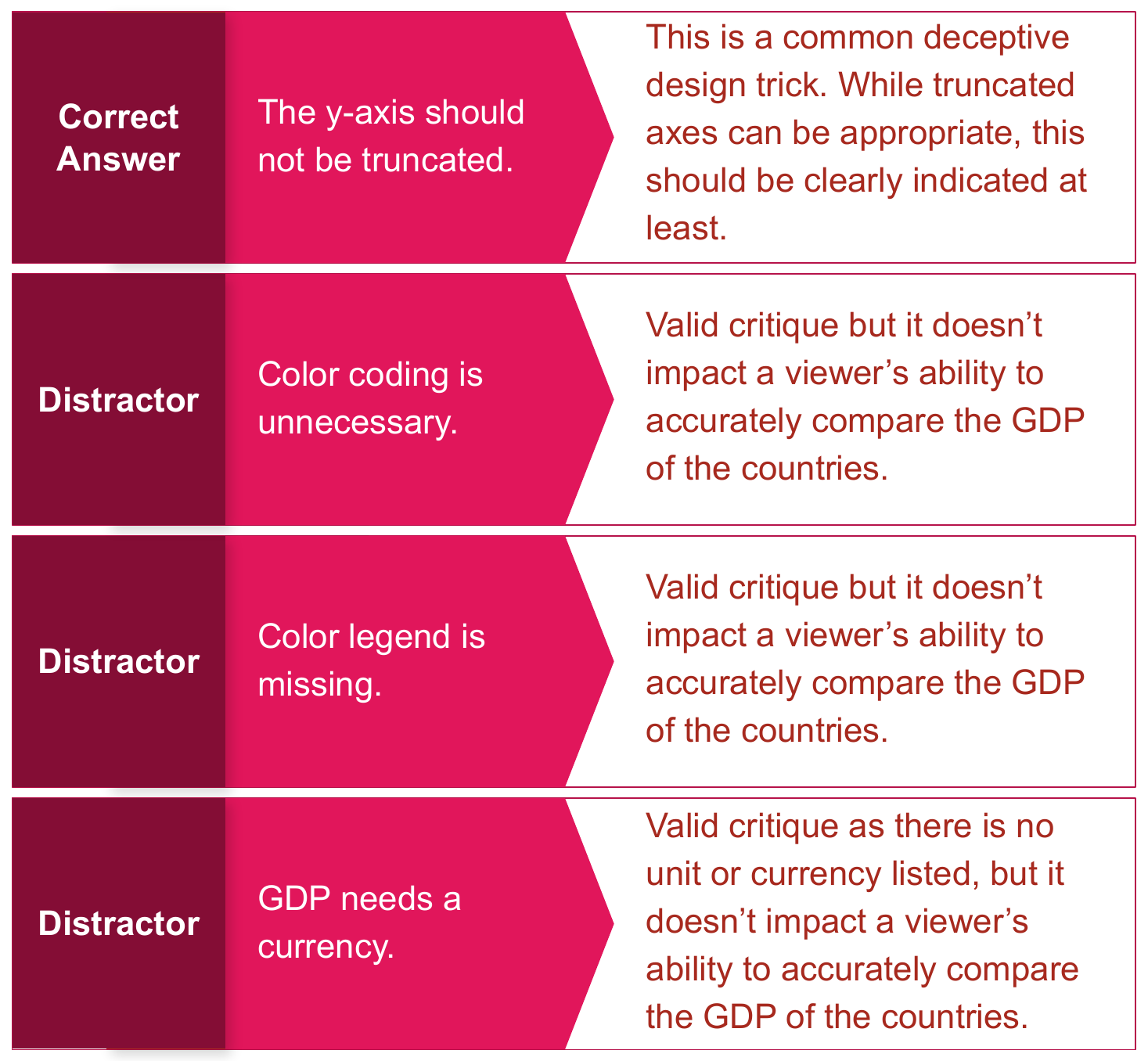}
    \caption{"Select only one of the following points to offer as feedback to the visualization creator that would most critically improve this visualization so that a person can accurately compare the GDP of the two economies."}
    \label{fig:distractorsEx1}
  \end{subfigure}%
  \hfill
  \begin{subfigure}{0.325\textwidth}
    \includegraphics[alt={A breakdown of answer options for the question in Q14 (Figure 4). Option A: Color Hue is more effective than Shape. This is the correct answer. Both are appropriate for nominal data, and color hue is the most effective for this plot according to Mackinlay’s order of effectiveness. Option B: Size is more effective than Color Hue. This is a distractor. Incorrect order. Size is not an appropriate encoding for nominal data, but color hue is appropriate. Option C: Color Saturation is more effective than Color Hue. This is a distractor. Incorrect order. Color saturation is not an appropriate encoding for nominal data, but color hue is appropriate. Option D: Shape is more effective than Size. This is a distractor. Correct order but not the most effective options. Size is not an appropriate encoding for nominal data but shape is appropriate.}, width=\linewidth]{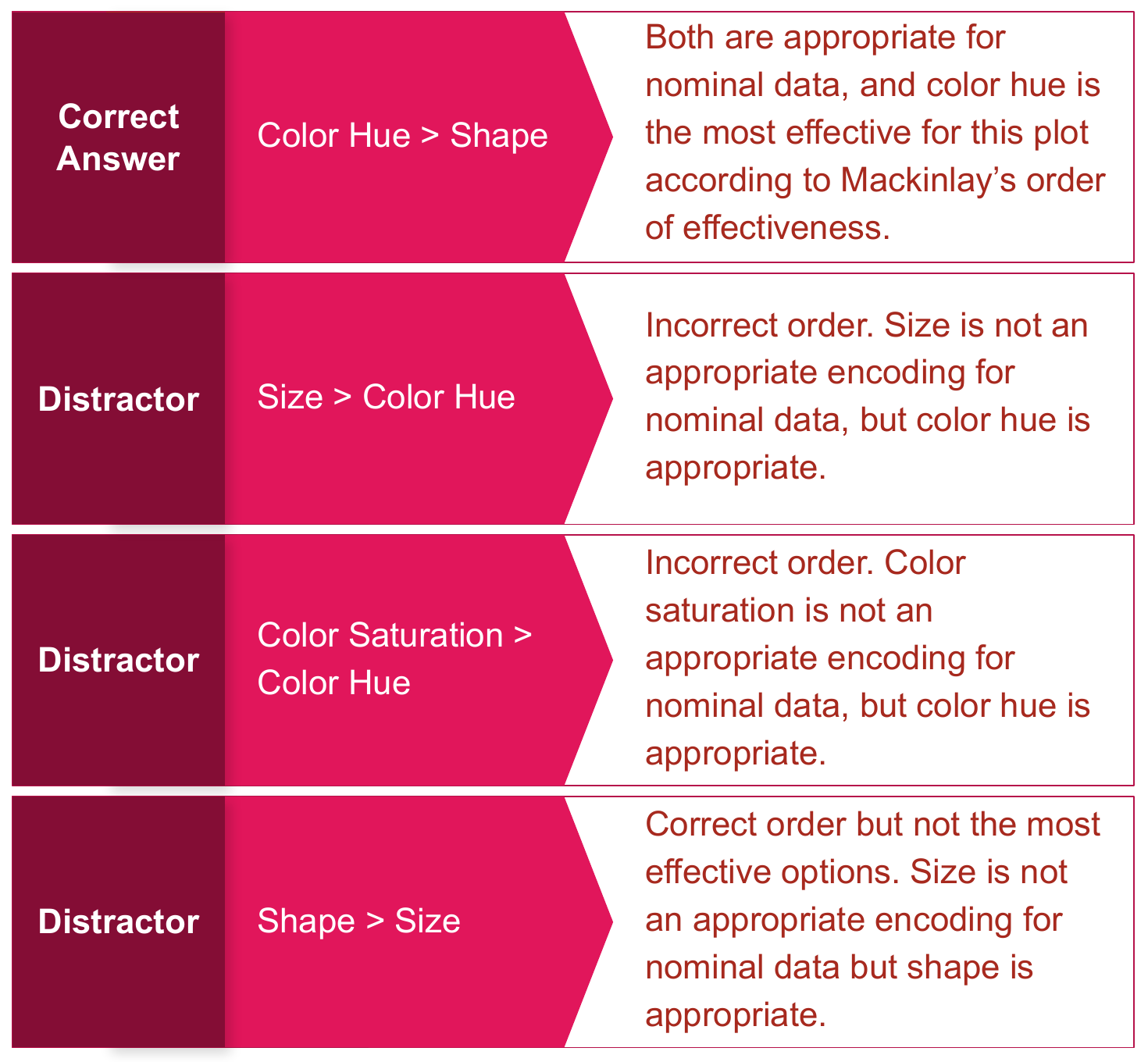}
    \caption{We want to add the region variable to our current scatterplot. What is the correct order of effectiveness for encoding the region variable from most to least effective?}
    \label{fig:distractorsEx3}
  \end{subfigure}%
  \hfill
  \begin{subfigure}{0.325\textwidth}
    \includegraphics[alt={A breakdown of answer options for the question in Q5 (Figure 5). Option A: Line Chart. This is a distractor. A line chart is an inappropriate chart type for the available data but are commonly misused by novices in such situations. Option B: Bar Chart. This is the correct answer. A bar chart is Ideal for comparing the GDP values in this case as it allows for rapid comparison to find min and max values. Option C: Choropleth Map. This is a distractor. A map is a popular choice as geographical information is available, but it isn’t best for rapidly identifying min and max values. Option D: Scatter Plot. This is a distractor. A scatter plot requires encoding another variable, which is unnecessary since the visualization goal is limited to the GDP.}, width=\linewidth]{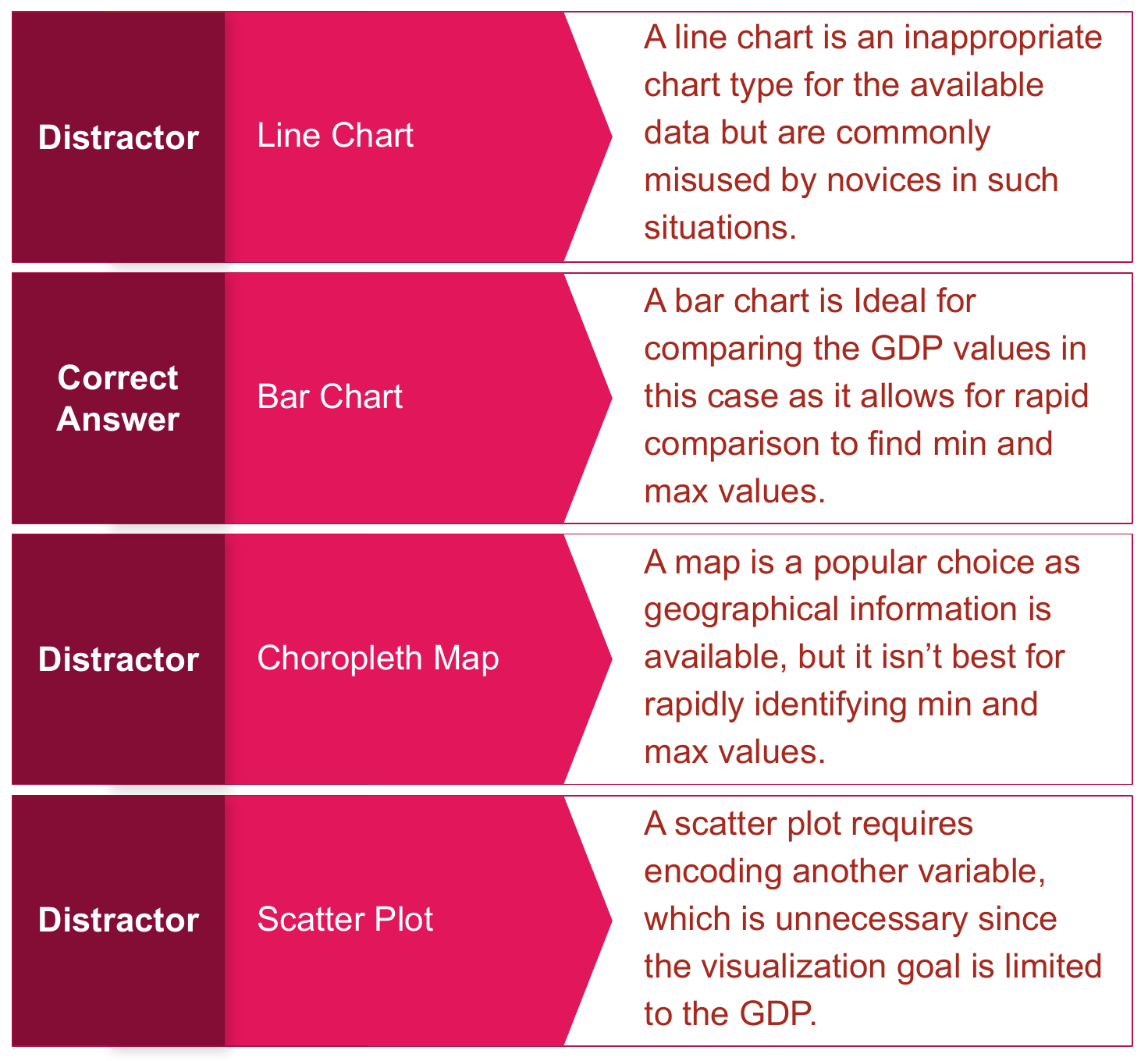}
    \caption{Visualization Design Goal: Help the viewer compare the GDP of these countries. Which chart type and rationale together make the most sense for comparing the values in the GDP column of the table above?}
    \label{fig:distractorsEx2}
  \end{subfigure}
  \caption{Explanation of the distractors and correct answers  for each of the three concept inventory questions depicted in this paper.}
\end{figure*}

\paragraph{Practicing Evaluation} When reviewing the literature for compelling application spaces, we found a prominent focus on the detection and mitigation of deceptive visualization tactics~\cite{pandey2015how}, including for visualization education~\cite{ge2023calvi,Bach2023VisEdChallenges}. Inspired by this work, Question 1 (Q1) emphasizes detecting and mitigating an established deceptive design tactic: truncating the y axis of a visualization~\cite{pandey2015how,ge2023calvi} (see \autoref{fig:example1-iteration}).
Versions one and seven of Q1 are depicted, showing how our iterative process led to significant improvements in the question design.
Beyond the y-axis truncation, the visualization also exhibits a number of distracting visualization design choices, including using a redundant color encoding. The distracting design choices are valid critiques of the visualization but fail to enhance the audience's ability to accurately compare the GDPs of these two countries. \autoref{fig:distractorsEx1} details the rationale behind the correct answer and the distractors.

\paragraph{Theory-driven Evaluation} This subcluster emphasizes the evaluation of visualizations using theoretical concepts.
As an example, Q3 emphasizes integrality versus separability of graphical marks, inspired by examples from Ware's influential textbook: Information Visualization: Perception for Design~\cite{ware2019information} (see \cite{rees2019survey} for a survey of relevant textbooks that informed our work). Specifically, Q3 asks students to rank visualizations by their alignment with the concept of integrality, testing their understanding of integrality and how they apply it to evaluate real visualization examples.

\paragraph{Redesigning Visualizations} To assess visualization redesign in a multiple choice format, we draw inspiration from graphical perception and visualization recommendation research. Specifically, we consider the idea of comparing pairs or sets of visualizations by their effectiveness for specific analysis tasks~\cite{moritz2019formalizing,zeng2023review}. Questions in this cluster present a fictional design scenario with a current visualization and design goal as well as 3-4 alternatives representing targeted design variations compared to the original. The student must select the best alternative that improves upon the original for the specified design goal(s).

\subsection{Understanding Visualization Design Principles}
\label{sec:concept-inventory:understanding-visualization-design-principles}

Given the abstract nature of the learning objectives in this subcluster, we focus on high level questions about \emph{remembering} and \emph{understanding} general ideas, concepts, and fields associated with visualization design. That being said, theories can be difficult to reason about in the abstract. Therefore, some concept inventory questions provide concrete scenarios that students can use to ground their understanding of the concepts.

\paragraph{Overview of Field} As an example, Q11 asks students to \emph{recall} subareas of visualization and \emph{understand} their main research contributions, which we assess using existing survey papers. For example, color research seems to emphasize human subjects experiments to measure perception of various color palette designs and effects~\cite{zhou2016survey,quadri2022visualization,zeng2023review} whereas network visualization research seems to emphasize algorithmic approaches for generating network visualizations~\cite{nobre2019state,filipov2023we}.

\paragraph{Understanding Theory} Assessment questions in this subcluster ask students to \emph{recall} and \emph{understand} specific theoretical concepts such as data types and rankings of encoding channels. We selected Mackinlay's rankings of encoding channels by task effectiveness~\cite{mackinlay1986automating} (inspired by work by Cleveland and McGill~\cite{cleveland1984graphical}) for designing these questions, which have been used widely in the visualization community~\cite{zeng2023review}. 
For instance, Q13 asks students to rank a given set of encoding channels by their theoretical effectiveness for a given data type.

\begin{figure}
    \centering
    \includegraphics[alt={Figure shows question 14 from the concept inventory. The question states ”Take a look at a sample below of the number of libraries in all 50 states in the USA.” The table has columns for “state”, “region”, “number of libraries”, and “population” and the states listed in the table are Ohio, Texas, New York, California, and Florida. Question continues to ask “Using the data about all 50 states, we create a scatter plot with population on the x-axis and number of libraries on the y-axis. We would like to answer the question: “Which region has the highest ratio of libraries per resident?” What is the correct order of effectiveness for encoding the region variable from most to least effective?“ Answer options are listed. Option A: Color Hue is more effective than Shape –Color hue allows for the eyes to easily visually cluster data points compared to shapes. Both color hue and shape are appropriate for nominal data. Option B: Size is more effective than Color Hue – Size differences are easy to spot for a limited number of regions whereas color hues can be difficult to differentiate. Size and color hue are appropriate for nominal data. Option C: Color Saturation is more effective than Color Hue – Color saturation will provide a more cohesive visualization design and more ideal for ordered data compared to color hue. Color saturation and color hue are appropriate for nominal data. Option D: Shape is more effective than Size – Different shapes reduce the likelihood of points being hidden compared to varying the size of points. Shape and size are appropriate for nominal data.}, width=\linewidth]{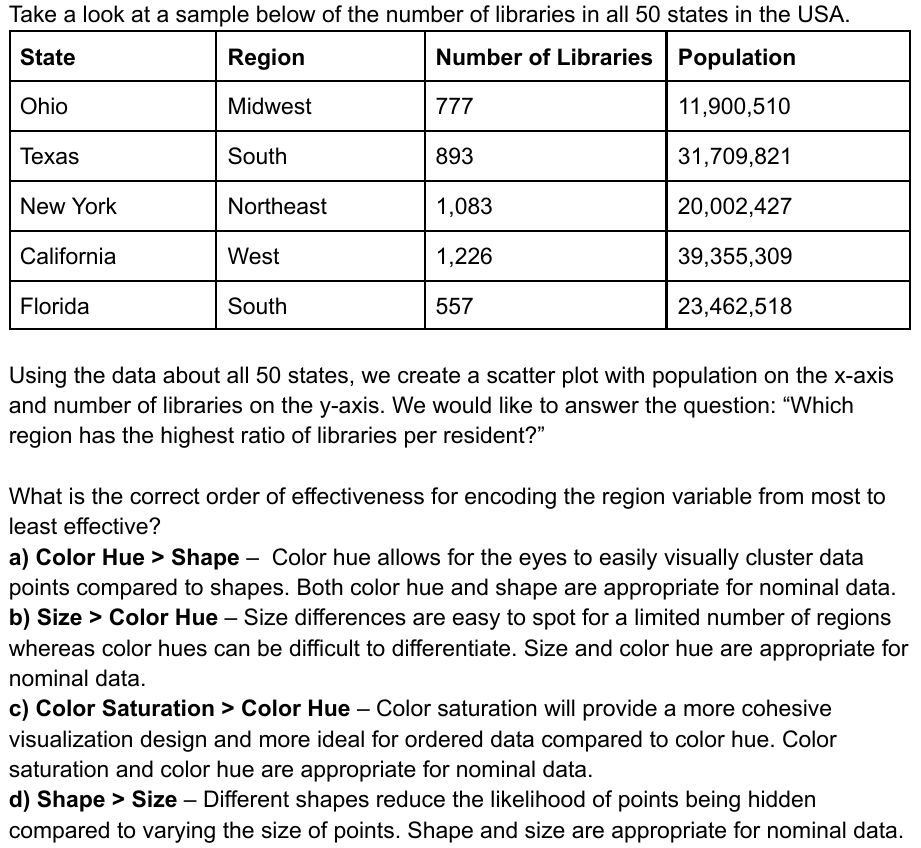}
    \caption{An example concept inventory question for the \textbf{Understanding Visualization Design Principles} cluster.} 
    \label{fig:example3CI}
\end{figure}

\paragraph{Understanding Methods} This subcluster assesses students' understanding of procedural methods rather than abstract theories, i.e., when and how to apply the theories. For instance, Q14 asks students to encode an additional variable in a scatterplot (see \autoref{fig:example3CI}). Students must decide which pair of encodings would be most effective and if they are correctly ordered in terms of effectiveness for nominal data.
In this case, a clear understanding of appropriate encodings per data type allows one to eliminate all but the correct answer. The distractors also force students to engage with encoding effectiveness rankings in a way they are not used to, i.e., shifting from picking the single best encoding to comparing pairs of encodings (see \autoref{fig:distractorsEx3}).

\subsection{Designing Visualizations}
\label{sec:concept-inventory:designing-visualizations}

This cluster emphasizes \emph{understanding} and \emph{applying} key concepts for visualization design, as well as \emph{creating} original visualizations.

\paragraph{Designing Effective Visualizations} This subcluster considers how students \emph{understand} and \emph{apply} their knowledge of marks, encoding channels, and effectiveness rankings to concrete datasets and analysis tasks. For example, Q5 gives students a data table and design goal written as an analysis task that the desired visualization must support. 
Specifically, Q5 asks students to select the most appropriate chart to compare the GDP of the 10 largest economies in the world (see \autoref{fig:example2CI}).
Students are given 4 chart descriptions along with design rationales, and students must select the best chart and rationale pairing for satisfying the design goal. Here, the distractors assess whether students can eliminate data columns and chart types that are irrelevant to the design goal (see \autoref{fig:distractorsEx2}).
For example, a line chart is not a valid solution for comparing the values of the GDP column, but it is included as a distractor as this is a common misuse of line charts that one of the authors witnesses in their own visualization course. The geographic emphasis of countries in the choropleth map represents another common design misconception.

\paragraph{Designing Original Visualizations} Given the limitations of multiple choice formats, the concept inventory does not allow students to generate new combinations of mark type and encoding channels (see \autoref{sec:discussion:lessons-learned} regarding limitations). That being said, students must understand the fundamental building blocks that can be composed before they can create novel visualization designs. Further, students must be able to reason about a large visualization design space as they create original visualizations and consider how existing design ideas may transfer to new visualization scenarios~\cite{bako2023understanding,hoque2020searching,parsons2026beyond}. Thus, the concept inventory questions in this subcluster measure how students \emph{understand}, \emph{analyze} and \emph{apply} these core building blocks as a prerequisite to creating original visualizations. For example, Q5 asks students to reason about chart descriptions rather than showing images, encouraging students to practice visualizing these charts in their minds to validate the given design choices and rationales.

\paragraph{Designing for Communication} Learning objectives in this cluster consider how visualizations are designed to communicate data to various audiences. For example, Q9 asks students to \emph{understand} and \emph{apply} visualization design techniques to communicate a specific takeaway message about an underlying dataset. Question design is informed by prior work on framing the visualization design problem around a key question or takeaway~\cite{stokes2025goodidea} or learning objective~\cite{lee-robbins2022learning,lee-robbins2023affective,adar2021communicative} to improve visualization design workflows.

\paragraph{Understanding the Design Process} This subcluster asks students to \emph{remember}, \emph{understand} and \emph{apply} their knowledge of common steps to visualization design workflows. However, \emph{applying} workflow knowledge may require computationally intensive activities (e.g., data profiling, data wrangling), so we focus on \emph{analyzing} high level workflow steps instead to reduce question duration and difficulty, informed by existing studies of workflow design (e.g., \cite{crisan2021passing,syeda2020designstudylite}). For instance, Q8 displays four ordered lists of common visualization design steps, each representing a partial workflow. Students must identify the answer that most closely resembles a realistic visualization design workflow.

%% file: content/6-discussion.tex
\section{Discussion and future work}
\label{sec:discussion}

In this work, we analyzed syllabi from 38 visualization courses and identify three learning objective clusters representing the common topics and skills associated with visualization design.
We developed a 14-question concept inventory demonstrating how these skills can be assessed in a multiple choice format.
In this section, we compare the concept inventory to existing visualization literacy assessments, discuss limitations  and lessons learned,
and describe next steps and
open research challenges revealed through our work.

\begin{figure}[]
    \centering
    \includegraphics[alt={Figure shows question 5 from the concept inventory. The question states ”Below is the forecasted 2026 GDP for the 10 largest economies in the world in millions of USD.” The table has columns for “country”, “continent”, “estimated 2026 GDP”, and “population” and the countries listed in the table are USA, China, Germany, Japan, India, UK, France, Italy, Canada, and Brazil. The stated visualization design goal is to “Help the viewer compare the GDP of these countries. Highest and lowest GDP should be rapidly identifiable in the chosen visualization.” Question continues to ask, “Which chart type and rationale together make the most sense for comparing the values in the GDP column of the table above?” Answer options are listed. Option A: 	Line chart, encoding Country on the x-axis and GDP on the y-axis, to make it easier to compare pairs of countries by connecting them together. Option B: Bar chart, encoding Country on the x-axis and GDP on the y-axis, to emphasize GDP differences using position and bar height. Option C: Choropleth Map, encoding GDP as color of the country on the map, to enable regional comparisons and to emphasize GDP differences using colors. Option D: Scatterplot, encoding Population on the x-axis and GDP on the y-axis with each point labeled with the Country, to make it easier for viewers to focus on the values of GDP and population, and not be distracted by the areas of bars/countries.}, width=\linewidth]{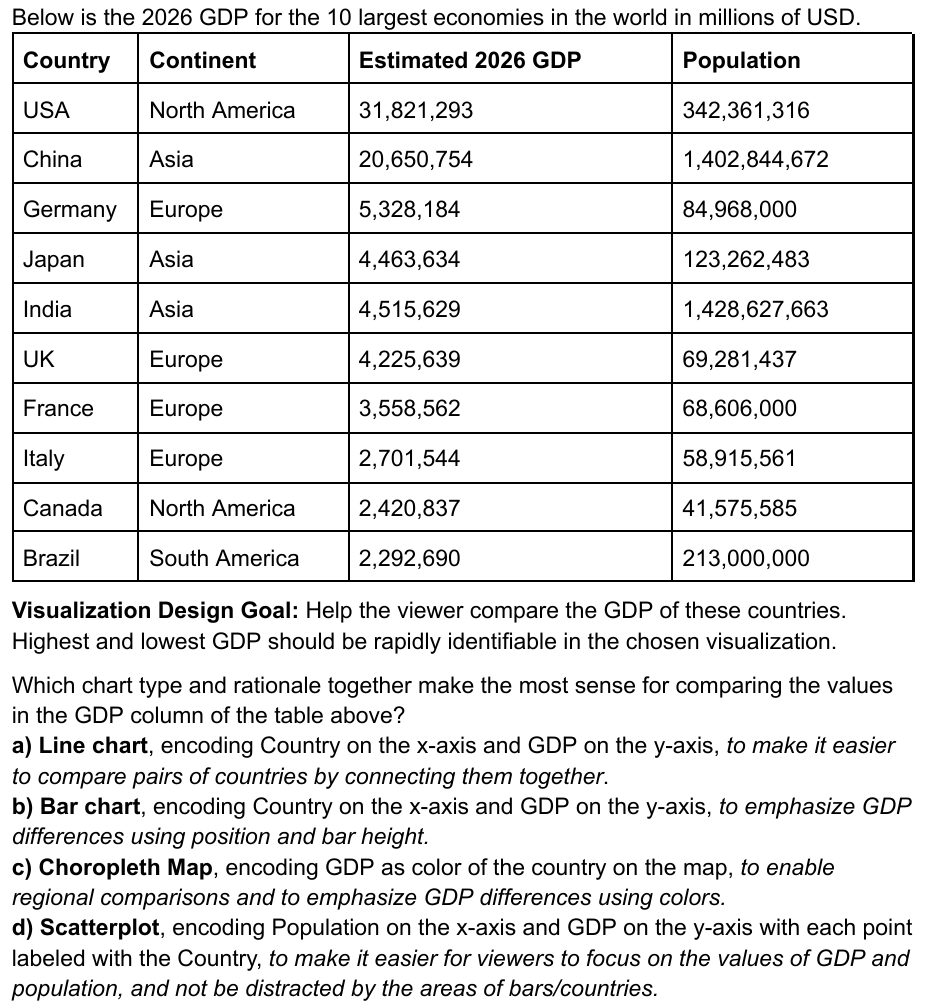}
    \caption{An example concept inventory question for the \textbf{Designing Effective Visualizations} subcluster.}
    \label{fig:example2CI}
\end{figure}

\subsection{Comparing With Existing Visualization Assessments}
\label{sec:discussion:CIversusOthers}

How does the concept inventory differ from existing visualization literacy assessments?
To answer this question, we compare the concept inventory with four existing assessments: CALVI~\cite{ge2023calvi}, AVEC~\cite{ge2025avec}, and VLAT/Mini-VLAT\cite{vlat,minivlat}.
We reviewed each assessment, for the skills they emphasize. Each skill is depicted as a separate row in \autoref{tab:CIcomparison}, and each skill paraphrases a specific design goal described in the corresponding paper(s). We add a checkmark for each assessment if it contains at least one question assessing this skill.
Then, we included additional rows for skills covered by our concept inventory but not the other assessments.
We observe significant overlap between these visualization literacy assessments and the concept inventory that we have developed. However, we also find that there are visualization tasks that are unique to the concept inventory which have not been previously assessed by the other assessments, suggesting that our focus on instructors' teaching goals revealed visualization design skills that were previously overlooked by existing assessments (\dgone). 

Furthermore, the concept inventory offers a broad scope of assessment (per \dgtwo) which differs from the other visualization literacy assessments.
If the goal is to gain in-depth insight into specific skills (e.g., thinking critically about visualization design errors/manipulation),
an existing assessment would be more suitable (e.g., CALVI).
Further, the concept inventory's design does not allow students to construct their own visualizations, a skill that AVEC is designed to address~\cite{ge2025avec}.

Consequently, the goal and intention of administering these assessments needs to be carefully considered to select an appropriate assessment. While CALVI, AVEC, and VLAT/Mini-VLAT provide in-depth insight into a person's performance on specific visualization design and analysis tasks, the concept inventory can provide a broad assessment of these tasks along with additional dimensions of visualization design.

\begin{table*}[]
\centering
\caption{Comparison between the visualization design concept inventory and existing visualization literacy evaluations (CALVI, AVEC, VLAT, and Mini-VLAT), showing the types of skills measured by each assessment.}
\label{tab:CIcomparison}
\begin{tabular}{@{}lllll@{}}
\toprule
\textbf{Visualization Skill}                                                                  & \begin{tabular}[c]{@{}l@{}}VLAT/\\ Mini-VLAT\end{tabular} & CALVI                     & AVEC                      & \begin{tabular}[c]{@{}l@{}}Concept \\ Inventory\end{tabular} \\ \hline
Complete specific visual analysis tasks using a given chart                                   & \checkmark                                 & \checkmark &                           &                                                              \\
\rowcolor[HTML]{EFEFEF} 
Interpret data illustrated within a given chart                                               & \checkmark                                 & \checkmark & \checkmark & \checkmark                                    \\
Think critically about visualization design errors/manipulations                              &                                                           & \checkmark &                           & \checkmark                                    \\
\rowcolor[HTML]{EFEFEF} 
Compare alternative visual encodings for a given chart                                        &                                                           &                           & \checkmark & \checkmark                                    \\
Construct new visualizations                                                                  &                                                           &                           & \checkmark &                                                              \\
\rowcolor[HTML]{EFEFEF} 
Compare chart and encoding selections using specific visualization design theories/principles &                                                           &                           &                           & \checkmark                                    \\
Recall common visualization design workflows                                                  &                                                           &                           &                           & \checkmark                                    \\
\rowcolor[HTML]{EFEFEF} 
Recall interdisciplinary influences outside of the visualization community                    &                                                           &                           &                           & \checkmark                                    \\ \bottomrule
\end{tabular}
\end{table*}

\subsection{Limitations and Lessons Learned}
\label{sec:discussion:lessons-learned}

\paragraph{Limitation: Pitfalls of Breadth over Depth.} The concept inventory prioritizes a broad assessment of visualization design skills over an in-depth assessment of specific skills. As a result, it is unable to provide in-depth assessment of individual skills.
Consequently, instructors may want to modify the concept inventory to cover specific theories and topics that they want to assess, or switch to an in-depth assessment that measures the desired skill. We discuss this further in \autoref{sec:discussion:future-research}. 

\paragraph{Limitation: North America and English Specific.} The concept inventory was evaluated only with instructors in North America and requires near-native English proficiency to assess visualization design skills. The context of the specific theories/principles, target audience, design goals, and visual encodings involved are described in English only.
Further, questions, distractors, and rationales can be nuanced in their wording, leading to potential misinterpretations by students who do not have the same cultural context and/or near-native proficiency. 
The language issue has been considered within prior work on concept inventories~\cite{forceCIHestenes} and can be addressed by translating the concept inventory to other languages and re-validating the translated version accordingly~\cite{FCIJapanese}. We see this as a viable path forward for the visualization design concept inventory as well. We posit that future work could also validate the concept inventory with students and instructors outside North America to test generalizability.

\paragraph{Limitation: Multiple Choice Questions.} Concept inventories are traditionally designed with multiple choice questions for various reasons (see \autoref{sec:related-work:concept-inventories}). While most skills can be evaluated in this format, some prove difficult to evaluate such as visualization construction (see \autoref{sec:discussion:CIversusOthers}). Optimal mixing of question types is still an open discussion in both visualization assessment and concept inventory research, which could be a promising direction for future work.

\paragraph{Lesson Learned: Decompose Compound Learning Objectives.}
Our clustering of learning objectives highlights how, despite differences in terminology or structures to specify learning objectives,
there appears to be a shared understanding among instructors of the core skills and concepts associated with visualization design.
That being said, one insight from our work that may benefit instructors is the idea of decomposing compound learning objectives. Instead of collapsing two or even three or four learning objectives together in one line, having a longer list of learning objectives may make it easier to assess whether a student is meeting a specific learning objective or not.

\paragraph{Lesson Learned: Design Context Matters.} For any given visualization problem, there are many viable solutions.
Feedback from interviewees was crucial to testing terminology and reducing ambiguity. For example, early versions of the questions lacked concrete details regarding the specific visualization design goal(s) that students should have in mind while selecting a correct answer, as well as crucial scenario-specific information, such as whether to consider color blindness and the visualization literacy of the target audience. P7 and P8 heavily emphasized the need to clearly define the visualization goal and target audience for each question as these two factors impact how visualizations are designed and critiqued and by extension which answer would be correct. In later interviews, P2, P4, P9, and P10 focused heavily on the wording used in the question and answer options, explaining how certain wording can be confusing or create ambiguity and certain terminology is not standardized (ex: categorical vs. nominal). Based on these experiences, we believe that instructors provide a distinct viewpoint for evaluating question context, which may prove useful for developing and evaluating future assessments.

\paragraph{Lesson Learned: How Should Developing Visualizations Be Assessed?} Despite visualization design being heavily utilized in the development of a visualization, questions related to this cluster were culled due to mixed opinions from interviewees on its relevance and importance for visualization design. Interviewees expressed concerns that concept inventory style questions were not appropriate for assessing a student’s ability to program visualizations (P2, P6, P8). There were major differences in how creating visualizations was covered in their courses with P1 not teaching specific visualization tools, P6  requiring a substantial project to be developed, P8 allowing vibe coding, and P10 allowing the use of AI-powered programming tools. Our findings reveal an underspecified component of visualization pedagogy. Future research could investigate the relationship between visualization design and visualization development and how (or whether) programming skills should be assessed within visualization courses.

\subsection{Evaluating the Visualization Design Concept Inventory}
\label{sec:discussion:next-steps}

Following the general approach of previous concept inventory designers (see \autoref{sec:related-work:concept-inventories}), we plan to conduct an empirically grounded validation study for our concept inventory as the next phase of our work. Drawing from the validation strategy outlined for FCS1~\cite{TewCSCI,6562691}, we plan to follow a mixed-methods approach that involves analyzing think-aloud interviews with students as well as their responses to our questionnaire. Think-aloud interviews can represent qualitative evidence of students' conceptual understanding and knowledge. Student responses to the concept inventory questionnaire can help refine and validate the instrument---for example, with the help of Item Response Theory~\cite{de2013theory, 10.1145/3287324.3287370} (a technique that can help determine the quality of individual questions), or by correlating overall student performance with their performance in external assessments (e.g., in a visualization design course, or with existing instruments such as AVEC~\cite{ge2025avec}). Furthermore, we hope that the results presented in this work will expand opportunities for discussion and feedback from the broader visualization education community, which will help us further refine the concept inventory during its validation phase.

\subsection{Using the Visualization Design Concept Inventory in Research and Practice}
\label{sec:discussion:future-research}

A concept inventory for visualization design can be useful in a range of contexts, especially once it is validated. Drawing inspiration from the recommended use of concept inventories as a ``diagnostic tool'' and ``for evaluating instruction''~\cite{forceCIHestenes}, we can see a concept inventory for visualization design being useful both inside and outside the classroom. 

Outside the classroom, the instrument can serve as assessments for individual learners in professional contexts or to evaluate professional development programs outside of traditional academic settings. Inside the classroom, in addition to assessment,
as suggested by Hestenes et al.~in their FCI paper~\cite{forceCIHestenes}, the concept inventory can be used to design class discussions to facilitate conceptual change---while the specific examples or datasets in these discussions can be discipline or context-specific, the overall contour of the discussion can be set by the questionnaire in the concept inventory, thus contributing to pedagogical design. Additionally, concept inventories have been used to assess the efficacy of teaching methods and changes \cite{ Sands2018CIsForUnderstanding} by administering the concept inventory to students before and after the course. The concept inventory we developed for visualization design could be used similarly. Furthermore, visualization instructors could use the concept inventory questions as a template to assess the efficacy of teaching theories and concepts not covered by the original concept inventory, by adapting the concept inventory to suit their needs as suggested with the FCI\cite{forceCIHestenes}.

We also feel that our work points to the possibility of concept inventories for more specialized areas of visualization design. For example, from the clusters shown in ~\autoref{tab:clusterstable}, visualization design for specific data domains can be represented by an additional set of concepts that builds on the concept inventory that we propose in this paper. 

Finally, we observe opportunities to expand on the concept inventory in the direction of assessment \emph{generation}, for example, to make it easier for instructors to create new exams and assignments for visualization courses. One approach is to generalize our concept inventory questions into reusable templates that can form the base for new, generated questions. Similarly, one could extend methods from existing visualization assessments to allow for automated generation of assessment questions similar to those in our concept inventory (e.g., extend AVEC~\cite{ge2025avec}).
Another exciting direction is to also consider how assessment questions may be generated using methods from computing education, such as Parsons problems for generating programming assessments~\cite{ericson2022parsons}. This would allow visualization researchers and educators to more easily incorporate development-focused assessments alongside the design-focused assessment questions in our concept inventory.

%% file: content/7-conclusion.tex
\section{Conclusion}

In this paper, we draw on existing methodology from the field of education to create a concept inventory for visualization design. The concept inventory was developed using learning objectives from 38 visualization course syllabi and refined through interviews with 10 visualization instructors, culminating in a 14 question multiple-choice assessment that covers a breadth of the most common visualization design skills that instructors emphasize in their courses.
This concept inventory creates new avenues of research in visualization education, for example, by revealing opportunities for developing new visualization assessments and by revealing open challenges in assessment design (e.g., how to assess visualization programming/development, how to introduce new question types into future assessments).
It can also serve as a tool for instructors to evaluate the effectiveness of their course and changes in instruction by using it as-is or using it as a template they can update to assess different visualization skills, theories, and methods.